\documentclass[superscriptaddress,aps,preprint]{revtex4-2}

\usepackage{graphicx}
\usepackage{dcolumn}
\usepackage{bm}
\usepackage{hyperref}
\usepackage{ulem}
\usepackage{verbatim}
\usepackage{amsmath}
\usepackage{times}

\begin{document}

\title{Real-space BCS-BEC crossover in FeSe monolayer}

\affiliation{State Key Laboratory of Low-Dimensional Quantum Physics, Department of Physics, Tsinghua University, Beijing 100084, China}
\affiliation{Department of Physics and Astronomy, University of Southern California, Los Angeles, CA 90089-0484, USA}
\affiliation{Jacobs University, Campus Ring 1, 28759 Bremen, Germany}
\affiliation{Division of Advanced Materials Science, POSTECH, San 31, Hyoja-dong, Nam-gu, Pohang 790-784, South Korea}
\affiliation{Frontier Science Center for Quantum Information, Beijing 100084, China}

\author{Haicheng Lin}
\thanks{These authors contributed equally to this work.}
\affiliation{State Key Laboratory of Low-Dimensional Quantum Physics, Department of Physics, Tsinghua University, Beijing 100084, China}

\author{Wantong Huang}
\thanks{These authors contributed equally to this work.}
\affiliation{State Key Laboratory of Low-Dimensional Quantum Physics, Department of Physics, Tsinghua University, Beijing 100084, China}

\author{Gautam Rai}
\affiliation{Department of Physics and Astronomy, University of Southern California, Los Angeles, CA 90089-0484, USA}

\author{Yuguo Yin}
\affiliation{State Key Laboratory of Low-Dimensional Quantum Physics, Department of Physics, Tsinghua University, Beijing 100084, China}

\author{Lianyi He}
\affiliation{State Key Laboratory of Low-Dimensional Quantum Physics, Department of Physics, Tsinghua University, Beijing 100084, China}

\author{Qi-Kun Xue}
\affiliation{State Key Laboratory of Low-Dimensional Quantum Physics, Department of Physics, Tsinghua University, Beijing 100084, China}
\affiliation{Frontier Science Center for Quantum Information, Beijing 100084, China}
\affiliation{Collaborative Innovation Center of Quantum Matter, Beijing 100084, China}

\author{Stephan Haas}
\affiliation{Department of Physics and Astronomy, University of Southern California, Los Angeles, CA 90089-0484, USA}
\affiliation{Jacobs University, Campus Ring 1, 28759 Bremen, Germany}

\author{Stefan Kettemann}
\affiliation{Jacobs University, Campus Ring 1, 28759 Bremen, Germany}
\affiliation{Division of Advanced Materials Science, POSTECH, San 31, Hyoja-dong, Nam-gu, Pohang 790-784, South Korea}

\author{Xi Chen}
\email{xc@mail.tsinghua.edu.cn}
\affiliation{State Key Laboratory of Low-Dimensional Quantum Physics, Department of Physics, Tsinghua University, Beijing 100084, China}
\affiliation{Frontier Science Center for Quantum Information, Beijing 100084, China}

\author{Shuai-Hua Ji}
\email{shji@mail.tsinghua.edu.cn}
\affiliation{State Key Laboratory of Low-Dimensional Quantum Physics, Department of Physics, Tsinghua University, Beijing 100084, China}
\affiliation{Frontier Science Center for Quantum Information, Beijing 100084, China}
\affiliation{Collaborative Innovation Center of Quantum Matter, Beijing 100084, China}

\date{\today}

\begin{abstract}
The quantum many body states in the BCS-BEC crossover regime are of long-lasting interest. Here we report direct spectroscopic evidence of  BCS-BEC crossover in real-space in a FeSe monolayer thin film by using spatially resolved scanning tunneling spectra. The crossover is driven by the shift of band structure relative to the Fermi level. The theoretical calculation based on a two-band model qualitatively reproduces  the measured spectra in the whole crossover range. In addition, the Zeeman splitting of the quasi-particle states is found to be consistent with the characteristics of a condensate. Our work paves the way to study the exotic states of BCS-BEC crossover in a two-dimensional crystalline material at the atomic scale.
\end{abstract}

\maketitle

Bardeen-Cooper-Schrieffer (BCS) superconductivity and Bose-Einstein condensation (BEC) are two asymptotic limits of a  fermionic superfluid. In the BCS limit, the Cooper pairs strongly overlap in space, and the condensation occurs at the onset temperature $T_{\rm{c}}$ of pair formation. On the other hand, the fermions in the BEC regime are bound into dimers  at a pairing temperature $T^*$ and the preformed dimers condense at a lower $T_{\rm{c}}<T^*$ as composite bosons. At the BCS-BEC crossover\cite{Leggett80,Nozieres85,Randeria14,Bloch08,Giorgini08,Chen05}, 
the coherence length (pair size) $\xi$ becomes comparable to the inter-particle distance ($\sim1/k_\text{F}$, where $k_\text{F}$ is the Fermi wave vector), or equivalently $\Delta/E_\text{F}\sim1$, where $\Delta$ and $E_\text{F}$ are the superconducting gap and the Fermi energy. The BCS and BEC limits are characterized by $k_\text{F}\xi\gg1$ and $k_\text{F}\xi\ll1$, respectively.

BCS-BEC crossover has been experimentally observed in  ultra-cold Fermi gases \cite{Jin03,Jochim03,Ketterle03,Salomon04,Grimm04,Jin04,Ketterle04,Thomas04}, where the inter-atomic interaction can be tuned by a magnetic field via the Feshbach resonance.  The condensed matter community is also highly interested in BCS-BEC crossover because of its possible relationship with high temperature superconductivity\cite{Chen05}. However, it is challenging to realize BCS-BEC crossovers  in  crystalline solid state systems because of the intricacy of tuning  the $\Delta/E_\text{F}$ ratio, as suggested in Refs.\cite{Eagles69,Nozieres99,Niroula20}.
Recently, several solid state systems with low carrier density, such as bulk FeSe-based superconductors\cite{Kanigel12,Matsuda14,Shin14,Matsuda16,Kanigel17,Okazaki20}, magic-angle twisted trilayer graphene\cite{park21} and lithium-intercalated zirconium nitride chloride\cite{Nakagawa21},  have shown evidences of BCS-BEC crossover. Most of these works demonstrate the renormalized band structure by angle-resolved photoemission spectroscopy, which only probes the occupied states under the Fermi level. The measurement of the predicted asymmetric gap at the Fermi level in the single-particle spectrum \cite{Chen05} is still not available.  In this work, we present direct spectroscopic evidence of a real-space BCS-BEC crossover in FeSe monolayers, where the local work function of the substrate is used to control the electronic band position and thus the carrier density. High energy resolution scanning tunneling spectroscopy (STS) reveals both occupied and unoccupied states of the BCS-BEC crossover regime, in particular the asymmetric gap structure at the unitary regime where $\Delta/E_{\rm{F}}\sim1$. More interestingly, this gap feature remains even as the hole band is completely shifted below the Fermi level. Our two-band model calculation qualitatively reproduces the main spectroscopic features of the BCS-BEC crossover in FeSe monolayer. It points out that the electron-hole inter-band interaction plays the key role for paring and the downward shifting of the hole band drives the BCS-BEC crossover in FeSe monolayer.

The experiments were performed in a commercial ultra-high vacuum (1$\times10^{-10}$ torr) scanning tunneling microscope (STM) equipped with molecular beam epitaxy. The base temperature of STM head could reach 60 mK with a relative high effective electronic temperature of 280 mK in samples \cite{chen19}. The n-type 6H-SiC(0001) (nitrogen-doped, resistivity 0.02-0.2 $\Omega\cdot$cm) substrates have been prepared by repeated annealing cycles of 10 minutes at 1450 $^{\circ}$C followed by 10 minutes of annealing at 500 $^{\circ}$C to form trilayer graphene (TLG) dominated surface. Both Bernal (ABA) and rhombohedral (ABC) stacking orders coexist in TLG as illustrated by Fig. 1(a) (see more in \cite{SM}). High quality FeSe monolayer films were synthesized by molecular beam epitaxy\cite{Chen11,Xue12}. High-purity Fe (99.995$\%$) and Se (99.999$\%$) were co-deposited onto the substrate at $\sim$400 $^\circ$C. The growth of FeSe was carried out under Se-rich condition and monitored by {\it in situ} reflection high-energy electron diffraction. The growth rate was about two monolayers per hour. All the STM and STS experiments were conducted with polycrystalline Pt-Ir alloy tips, which were treated by poking on Ag(111) island surface until correct topographic and spectroscopic features were obtained. The d$I$/d$V$ spectra on FeSe films were acquired by the standard lock-in technique with a modulation frequency f=887 Hz.

\begin{figure*}[t]
		\setlength{\abovecaptionskip}{-0.8cm} 
       \includegraphics[width=6.5in]{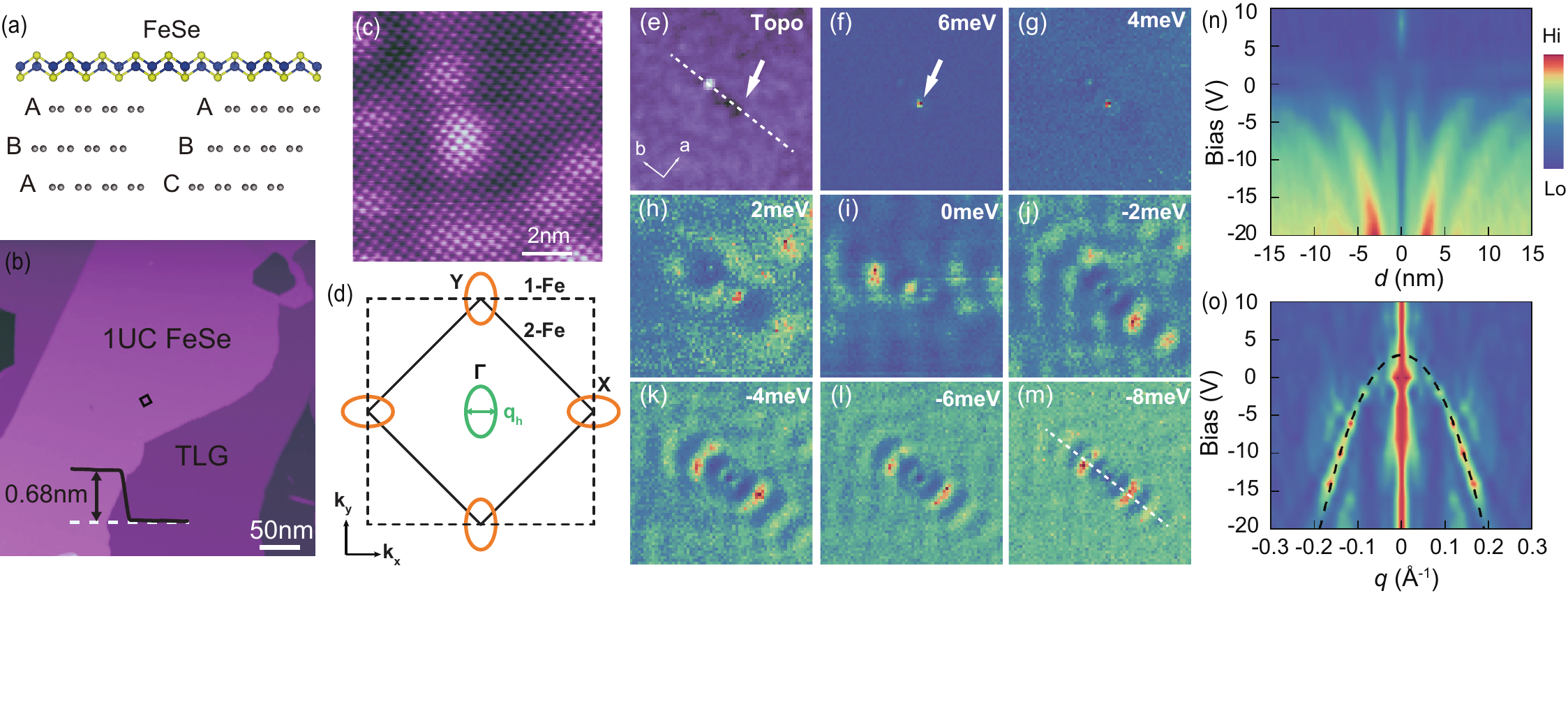}
       \caption[]{
The FeSe monolayer. (a) Side view of  monolayer FeSe on TLG. (b) Topography image of a 390 nm $\times$ 390 nm area of FeSe monolayer film on TLG/6H-SiC(0001) substrate (sample bias: 3 V, tunneling current: 20 pA, temperature: 280 mK). (c) Atomically resolved STM topography (10 nm $\times$ 10 nm, 0.02 V, 0.1 nA) acquired at the position marked by the black square in (c). (d) Schematic of the Brillouin zone and the Fermi surface of a FeSe monolayer.  (e) The topography image (35 nm $\times$ 35 nm, -20 mV, 0.1 nA, lock-in oscillation $V_{\rm{osc}}$ = 0.2 mV, 280 mK) of an  area for d$I$/d$V$ mapping. (f-m) d$I$/d$V$ mapping with energies from 6 meV to -8 meV. (n) Spatially resolved d$I$/d$V$ spectra along the white dashed line in (E). (o) Fourier transform of d$I$/d$V$ mapping with scattering vector along the direction of the white dashed line. 
}
\end{figure*}

Figure 1(b) shows a typical STM  image of the atomically flat FeSe monolayer on TLG.   No Moir\'{e} pattern has been observed indicating weak coupling between epitaxial FeSe monolayer and the substrate.  The apparent height of the FeSe monolayer is 0.68 nm [Fig. 1(b)], which is larger than the lattice constant 0.550 nm along the c-axis of bulk FeSe. The atomically resolved STM image  [Fig. 1(c)] reveals an in-plane lattice constant of 0.375 nm which is consistent with the 0.377 nm of bulk FeSe \cite{Wu08}. 

The 2D crystalline semi-metallic FeSe monolayer is an ideal platform to realize the BCS-BEC crossover because the Fermi energy is only a few meVs and can be fine-tuned by the underneath graphene layers\cite{Wantong21} to the size comparable to the superconducting gap. The electronic band structure of monolayer FeSe is similar to that of its bulk counterpart, consisting of hole pockets near the $\Gamma$ point and  electron pockets near the M points of the Brillouin zone [Fig. 1(d)]. The band dispersion of the hole pocket of FeSe monolayer is extracted from the STS mapping in the vicinity of defects. Figure 1(e) shows an area with two defects: an adatom (bright protrusion) and a vacancy. The vacancy scatters the electrons more strongly than the adatom. The local density of states (LDOS) modulated by the vacancy scattering is revealed by the d$I$/d$V$ mapping as a function of energy [Figs. 1(f-m) and see more details in Fig. S5]. Ripples of density of states are clearly observed if the bias voltage is below 4 mV.  Figure 1(n) summarizes the LDOS along the white dashed line (also the direction of b-axis) in Fig. 1(e). The monotonic increase of the wave length with energy manifests  hole-like behavior. The energy dispersion of the scattering vector of the hole pocket along the b-axis [Fig. 1(o)] is obtained via Fourier transform of the d$I$/d$V$ mappings, followed by a parabolic fitting with an effective mass of $1.4\pm0.1$ m$_0$, where m$_0$ is the free electron mass. The measured Fermi energy fluctuates spatially and is found to be $3.1\pm0.5$ meV for the area shown in Fig. 1(e).  

The tunneling current near the Fermi energy from a regular tip mainly consists of electrons with small lateral momentum. Due to the tunneling matrix effect, the d$I$/d$V$ spectra of FeSe monolayer are dominated by the hole pocket at the $\Gamma$ point \cite{Davis17}. To reveal the band dispersion of the electron pockets, an extremely sharp STM tip is needed to access the electronic states near the M point (Fig. S6).

\begin{figure*}
       \begin{center}
       \includegraphics[width=6.5in]{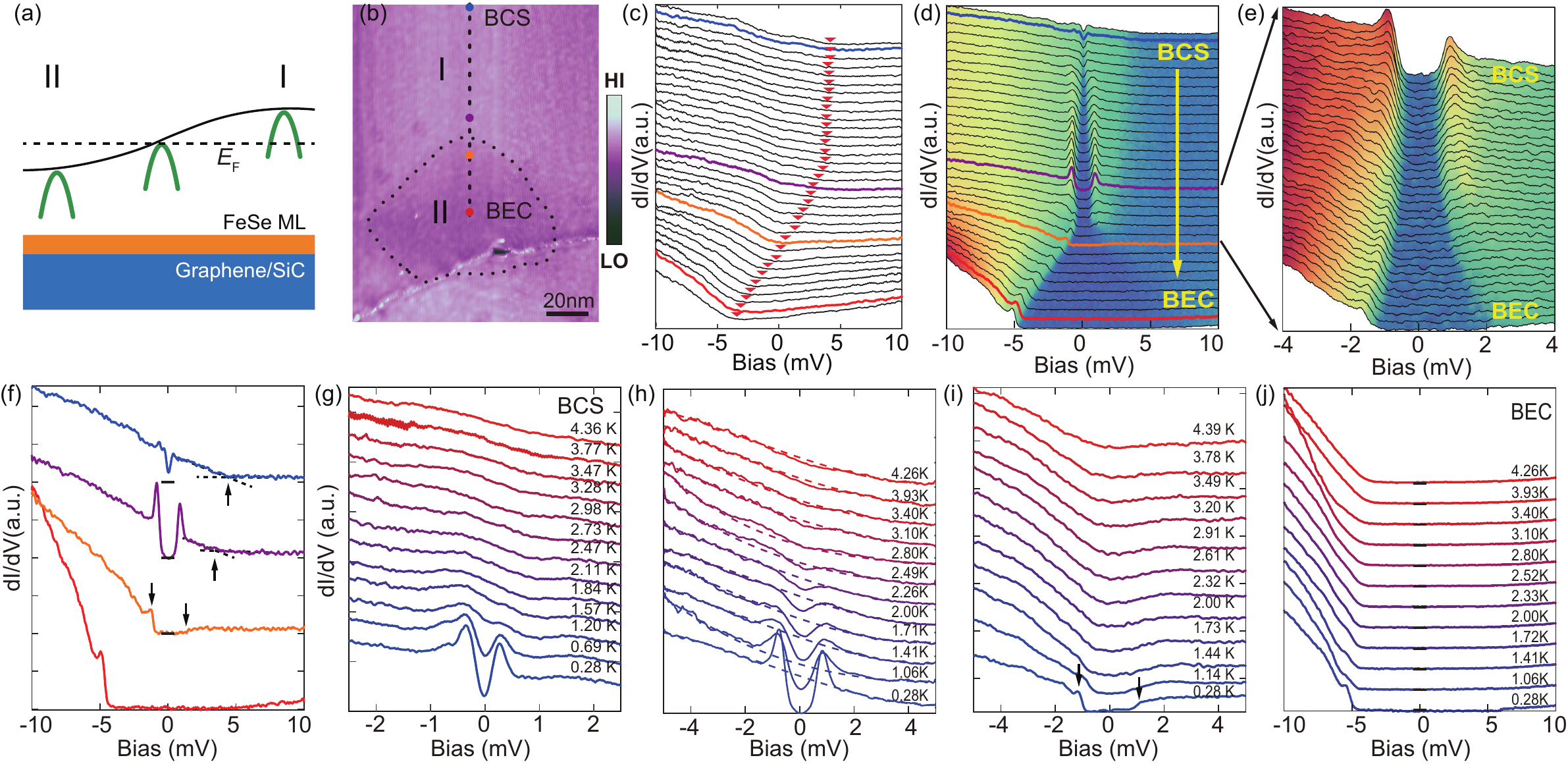}
       \end{center}
       \caption[]{
Real-space BCS-BEC crossover in monolayer FeSe. (a) The hole band shift relative to the Fermi level from region \uppercase\expandafter{\romannumeral1} to region \uppercase\expandafter{\romannumeral2}.(b) FeSe monolayer sheet on TLG (116 nm $\times$ 150 nm, -10 mV, 40 pA). The dashed line from the blue to the red dots is 93 nm. (c) Spatially resolved d$I$/d$V$ spectra (-10 mV, 0.1 nA, $V_{\rm{osc}}$ = 0.1 mV, 4.5 K) along the dashed line in (a). Vertical offsets have been added for clarity. (d) Spatially resolved d$I$/d$V$ spectra (-10 mV, 0.1 nA, $V_{\rm{osc}}$ = 0.1 mV, 0.28 K) along the dashed line in (a). Vertical offsets have been added for clarity. (e) The d$I$/d$V$ spectra with higher spatial resolution in the boundary region of (d).(f) Spectra (-10 mV, 0.1 nA,$V_{\rm{osc}}$ = 0.1 mV, 0.28 K) acquired at the locations marked by colored dots in (a). The arrows indicate the top of hole band. (g-j) Temperature dependent d$I$/d$V$ spectra of (f).[$V_{\rm{s}}$ = -3 mV, $I_{\rm{t}}$ = 0.1 nA, $V_{\rm{osc}}$ = 30 $\mu$V for (g). $V_{\rm{s}}$ = -5 mV, $I_{\rm{t}}$= 0.1 nA, $V_{\rm{osc}}$ = 50 $\mu$V for (h-j).]
}
\end{figure*} 

The spatial variation of the Fermi energy of electronic band of FeSe monolayer on top of TLG [Fig. 2(a)] can be directly probed by the spatially resolved d$I$/d$V$ spectra. Figure 2(b) shows a piece of FeSe monolayer extending over several hundred nanometers on TLG. A band shift is evident in the d$I$/d$V$ spectra at 4.5 K (Fig. 2(c), also see Fig. S7) along the dashed line in Fig. 2(b). The kinks indicated by the red triangles correspond to the top of the hole band, for example, 3.5 meV, 3.0 meV, 0.3 meV and -3.3 meV for the four dots in Fig. 2(b), respectively. The top of the hole band gradually drops along the dashed line. Eventually the hole band  becomes fully occupied [Fig. 2(a)], and only the electron pockets contain the Fermi surface. Such band shift is caused by the chemical potential alignment with the TLG substrate \cite{Wantong21}, where the local work function is spatially inhomogeneous\cite{SM}. 

With descending temperature, a phase transition occurs, as indicated by new features in the spectra of Figs. 2(d) and 2(e) along the same dashed line in Fig. 2(b). It shows the LDOS at various locations along the FeSe monolayer at the base temperature of 0.28 K. The distinction of the spectra of four locations indicated by four color dots in Fig. 2(b) is more clear as shown in Fig. 2(f). The d$I$/d$V$ spectra on region \uppercase\expandafter{\romannumeral1} and away from the border [blue dot in Fig. 2(b)] exhibit a superconducting gap of $\Delta=0.3$ meV [blue curve in Fig. 2(d,f)], similar to the previous study \cite{Wantong21}.  The temperature-dependent d$I$/d$V$ spectra of this region from 0.28 K to 4.36 K are shown in Fig. 2(g). The superconducting gap is completely closed at about 1.2 K. The ratio $2\Delta/k_{\rm{B}}T_{\rm{c}}$ is estimated to be about 5.8, which is larger than BCS ratio 3.53 indicating it is in the strong coupling regime. In this area, the ratio $\Delta/E_{\rm F}\sim$ 0.1 is already larger than that of the most unconventional superconductors in terms of the Uemura plot \cite{PhysRevMaterials.3.104802,singh2018superconducting}. 

As the tip moves toward the region \uppercase\expandafter{\romannumeral2} along the dashed line in Fig. 2(b), the top of the hole band drops as shown in Fig. 2(a). At the same time, the gap size increases (0.9 meV for the purple dot) and the coherence peaks become more pronounced. The sloped background in the spectrum originates from the LDOS of the hole pocket. Once the background slope is subtracted, the particle-hole symmetry of the gap is preserved. The temperature-dependent d$I$/d$V$ spectra in Fig. 2(h) shows that the gap-closing temperature reaches 2.8 K in this regime and the $2\Delta/k_{\rm{B}}T_{\rm{c}}$ ratio increases to about 7.5. The enhanced superconductivity could be explained by stronger electronic correlations at lower hole density \cite{PhysRevLett.97.107001,PhysRevB.98.064512} or the possible enhancement of superconducting pairing mediated by the electron pockets. In any case, it appears that a BCS-BEC crossover unitary regime has been  reached at the interface. For example, the $\Delta/E_{\rm F}$ ratio is as high as 0.3 on the purple dot.  

When the tip moves to the orange point in Fig. 2(b), the d$I$/d$V$ spectrum [orange curve in Fig. 2(f)] shows the asymmetric step-like gap structure predicted for BCS-BEC crossover\cite{Chen05}. The energies of the two gap edges indicated by arrows on the orange curve are symmetric relative to the Fermi level while the  heights are highly asymmetric. No sharp coherence peaks present in the orange curve. As the temperature increases [Fig. 2(i)], the step-like gap structure disappears at around 2 K. However the depression of LDOS around the Fermi level persists to higher temperature. Such behavior resembles the predicted pseudogap in the unitary regime of BCS-BEC crossover. More experiments are needed to further confirm the pseudogap state in this regime.

As the tip moves further and across the border [on region \uppercase\expandafter{\romannumeral2} in Fig. 2(b)], the gap size keeps increasing [see red curve in Fig. 2(d)] and the shape becomes highly asymmetric: the gap edge at the negative bias is still distinct whereas the corresponding feature at the positive bias gradually diminishes. In addition, the asymmetry cannot be eliminated by background subtraction. The continuous evolution of the LDOS is more clearly demonstrated by the spatially resolved d$I$/d$V$ spectra [Figs. 2(e) and 2(f); see more in Fig. S8\cite{SM}]. In the lower part of Fig. 2(e), a feature similar to that of a superconducting gap emerges at the top edge of the hole band, even although the band is already fully occupied.  This distinct feature of the band edge is absent when the top of hole band is above the Fermi level as in region \uppercase\expandafter{\romannumeral1}, implying that such feature does not come from a regular band edge.  The variable temperature experiment shows that this weak peak feature fades away when the temperature increases up to around 2 K [Fig. 2(j)]. Furthermore the absence of hole band on the Fermi level helps the electron pockets to stand out in the LDOS with a sharp STM tip. The small dip in the d$I$/d$V$ spectrum (Fig. S9) indicates weak superconductivity of the electron pockets.

We attribute above evolution of the spectra across the dotted line border in Fig. 2(b) to a BCS-BEC crossover. The BCS and BEC regimes are labelled in Fig. 2. In the BEC regime, the paired electrons in the hole pocket undergo Bose-Einstein condensation. 


In order to further investigate this BCS-BEC crossover behavior, we consider the two-band model defined in \cite{Chen15,Chubukov16} as a minimal model of 
the multiband superconducting model 
for the FeSe monolayer\cite{Bang2014,Koshelev14}. 
This model [sketched in Fig.~3(a)] consists of two parabolic bands: a hole-like band centered at the $\Gamma$~point with dispersion $\xi^h_k = E_V - k^2/2m_h$, and an electron-like band centered at momentum $Q$ with dispersion $\xi^e_k = E_C +  (k-Q)^2/2m_e$. $E_V$ and $E_C$ are the respective band edges of the hole and electron bands, and  $m_h$ and $m_e$ are the corresponding effective masses. In FeSe, the two bands overlap in energy,
so that the  parameter $\Delta_0 = (E_C-E_V)/2 <0$ is negative. 

We assume that the dominant pairing interaction is spin fluctuation mediated inter-band scattering with interaction strength $U = -V_{SF}>0$\cite{Mazin08,Kuroki08}. The interaction is finite in an energy interval of width $2\Lambda$ around the chemical potential $\mu$. At zero temperature, BCS theory gives us self-consistency equations for the BCS order parameters in the two bands, $\Delta_e$ and $\Delta_h$,
\begin{eqnarray}\label{SelfConsistency}
\Delta_e= \Delta_h \frac{U}{2}\int_{\mu-\Lambda}^{\mu+\Lambda}d\xi_k\frac{\rho_h\left(\xi_k\right)}{\sqrt{\left(\xi_k-\mu\right)^{2}+\Delta_h^{2}}},
\nonumber \\
\Delta_h= \Delta_e \frac{U}{2}\int_{\mu-\Lambda}^{\mu+\Lambda}d\xi_k\frac{\rho_e\left(\xi_k\right)}{\sqrt{\left(\xi_k-\mu\right)^{2}+\Delta_e^{2}}},
\end{eqnarray}
where $\rho_h(\xi_k)$ ($\rho_e(\xi_k)$) is the densities of states in the hole-like (electron-like) band . The chemical potential $\mu$ is determined by particle number conservation,
\begin{eqnarray} \label{ParticleNumberConservation}
n &=& \int_{E_{\rm C}}^D d \xi_k  \rho_e \left(\xi_k\right) \left( 1- 
\frac{ \xi_k -\mu}{\sqrt{\left(\xi_k-\mu\right)^{2}+\Delta_e^{2}}} \right) \nonumber \\
&+& \int_{-D}^{E_{\rm V}} d \xi_k  \rho_h\left(\xi_k\right) \left( 1- 
\frac{ \xi_k -\mu}{\sqrt{\left(\xi_k-\mu\right)^{2}+\Delta_h^{2}}} \right).
\end{eqnarray}
We assume that the films have a constant 2D density of states, $\rho_e(\xi) = \frac{m_e}{2\pi}$ when $E_C<\xi<D+E_C$, and $\rho_h(\xi) = \frac{m_h}{2\pi}$ when $-D+E_V<\xi<E_V$, and that $m_e = m_h$. All pertinent features of the BCS-BEC crossover reported in this experiment can be reproduced with this simple choice, which has the advantage that the integrals in the formula (1) of the manuscript 
and the formula \eqref{ParticleNumberConservation} here can be performed exactly. 

The parameter regime applicable to the experiment is defined by the following limits: (1) the pairing gaps  are small compared to the range of the attraction: $\Lambda\gg\Delta_e, \Delta_h$; (2) the chemical potential is close to the band edges: $\Lambda \gg |\mu - E_C|, |\mu-E_V|$; and (3) the width of the bands is larger than the pairing gaps $D\pm\mu\gg\Delta_e, \Delta_h$.

In Fig.~3(b), we show the phase diagram of this model as a function of the hole doping parameter $\epsilon_V = E_V - \epsilon_F$ and the gap/overlap parameter $\Delta_0$. Both axes are plotted in units of the binding energy $\Delta_m = 2\Lambda \exp\left(-2/U\rho\right)$. In the semimetal regime, $\Delta_0<0$, there is always a non-zero order parameter $\Delta_{h}, \Delta_e>0$. The crossover between the BCS and BEC regimes occurs close to $\mu = E_V$. When the chemical potential is far above (below) the hole band edge, the hole band is in the BEC (BCS) regime. 

In order to model the real-world material, we choose parameters for the two-band model to reproduce the properties of FeSe monolayers. Within the limits described above, the two-band model is uniquely described by choosing the set of parameters $(\Delta_m, \Delta_0, \Lambda)$. The solutions of the equation are effectively independent of $\Lambda$ as long as it is chosen to be large enough. We use $\Lambda = 50$~meV. The band overlap parameter must be chosen such that the density of overlapping carriers between the hole and electron bands is comparable to real FeSe. Since we are approximating a density of states that scales as $\sqrt{\xi_k}$ by a constant density of states, the band overlap in the model has to be fixed at $2\Delta_0 = -\frac{|2\Delta_0^{real}|^{3/2}}{\Lambda^{1/2}}$, where $2\Delta_0^{real} = -5.8$~meV is the band overlap in real FeSe. Finally, we choose the binding energy $\Delta_m$ such that the pairing gap in the hole band in the BCS regime is comparable to the experimental value of $\Delta_h^{BCS} = 0.9$~meV. We empirically determine this to be $\Delta_m = 1.1$~meV.

We use the experimentally measured values of the doping parameter $\epsilon_V(x)$ across the dashed line boundary [shown by the red triangles in Fig.~2(b)] and calculate the self-consistent order parameters and chemical potential according to \eqref{SelfConsistency} and 
formula (1) in \cite{SM}. At each position, the quasiparticle density of states in the hole band can be calculated using the quasiparticle propagator,
\begin{align}\label{DOS}
    DOS_h(E) =& \sum_{\eta = \pm 1}\text{Re}\left[{\rho_h\left(\mu +\eta \text{sign}(E)\sqrt{(E^2 - \Delta_h^2)}\right)}\nonumber\right.\\& \left.{\times\left(\frac{|E|}{\sqrt{E^2 - \Delta_h^2}} -\eta\text{sign}(E) \right)}\right]
\end{align}
Fig.~3(d) shows the density of states calculated across the dashed line boundary [Fig. 2(a)] in the two-band model. Note the remarkable agreement with the experimental data in Fig.~2(e) despite the simplicity of the model. We needed to specify only two essential parameters, the binding energy $\Delta_m$ and the adapted overlap parameter $\Delta_0$. In the BEC regime, when the chemical potential is above the hole band edge, there is no singularity in the density of states, but it is still peaked, and the order parameter is finite. In the BCS regime, when the chemical potential is within the hole band, the density of states diverges when $E = \Delta_h$. To show this, we have plotted characteristic traces of the density of states from both regions in Fig.~3(c). 

The response to magnetic field is an essential property of a condensate and helps to identify the different phases. We also show how the density of states looks in the presence of an applied magnetic field. The expression for the density of states becomes a sum of the Zeeman-shifted spin-up and spin-down terms,
\begin{align}
    DOS_h(E;B) = \sum_{\sigma = \pm 1} \frac{1}{2}DOS_h(E+\sigma g\mu_B B;B=0),
\end{align}
where $\mu_B$ is the Bohr magneton, and $g$ is the Land\'{e} factor. The resulting density of states in the BEC regimes is shown in Fig. 3(e).

\begin{figure}
       \begin{center}
       \includegraphics[width=3.25in]{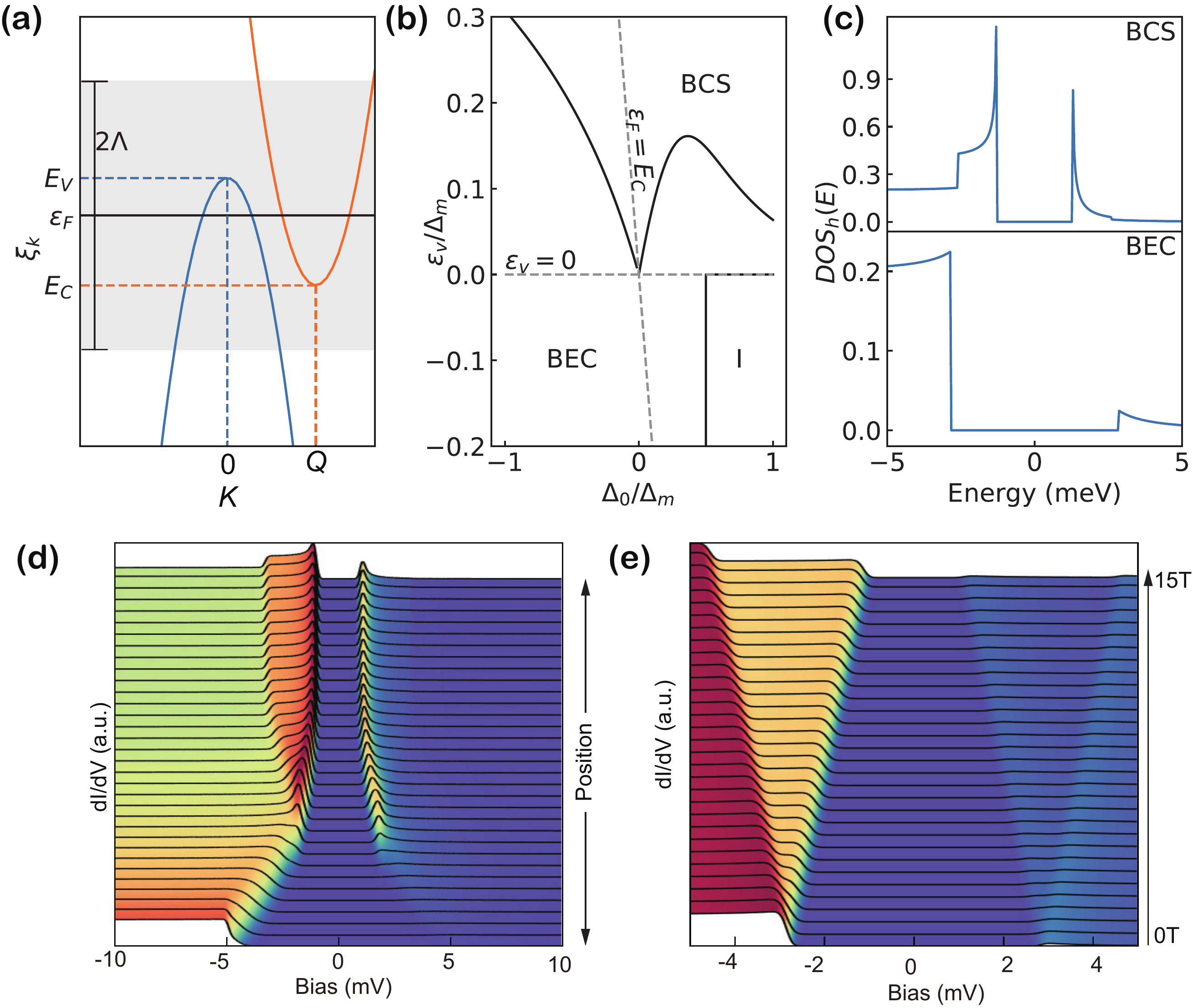}
       \end{center}
       \caption[]{
Two-band model for FeSe. (a) An electron-like band at momentum $Q$ overlaps with a hole-like band at zero momentum in the energy interval $[E_C, E_V]$. There is an attractive interband attraction within an energy range of $2\Lambda$ around the Fermi level. (b) Phase diagram of the two-band model depending on the hole doping parameter $\epsilon_V = E_V - \epsilon_F$ and the gap (or overlap) parameter parameter $\Delta_0 = (E_C-E_V)/2$. In FeSe, $\Delta_0<0$, leading to a a BCS-BEC crossover in the hole band as the hole band is depleted with decreasing $\epsilon_V$. (c) The hole density of states in the BCS and BEC regimes. In the BCS regime, the coherence peaks are at $\pm\Delta_h$. In the BEC regime, the coherence peaks are at $\pm \sqrt{\Delta_h^2 + \mu_h^2}$. (d) The density of states across the BCS-BEC interface using experimental values for the hole doping parameter along the dashed line in Fig.~2(a). (e) The density of states in the BEC regime as a function of an applied magnetic field. 
}
\end{figure}

\begin{figure}
       \begin{center}
       \includegraphics[width=3.25in]{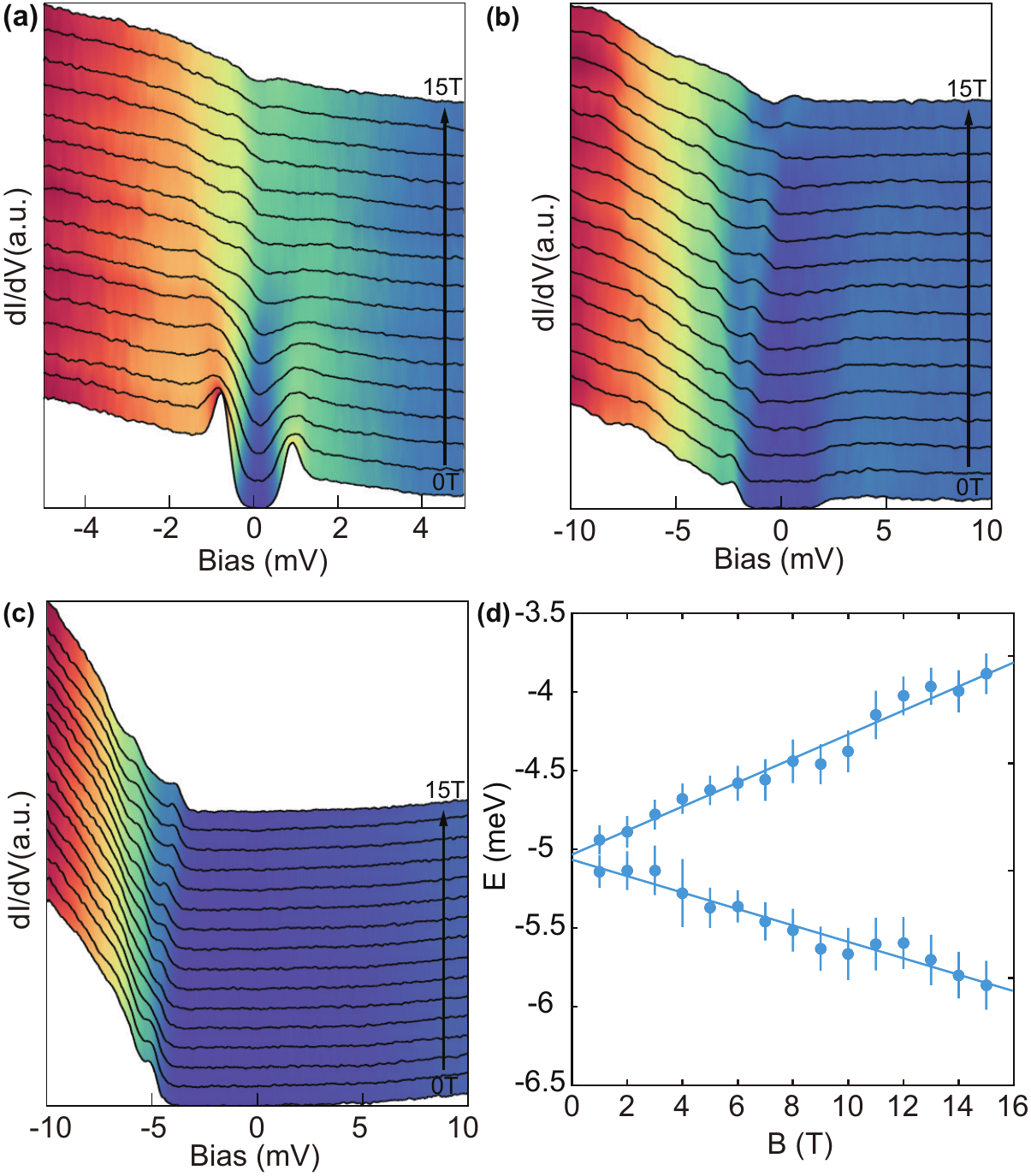}
       \end{center}
       \caption[]{
The d$I$/d$V$ spectra of FeSe monolayer under perpendicular magnetic field up to 15 T.  (a) Spectra in the BCS regime (-3 mV, 0.1 nA, $V_{\rm{osc}}$= 0.05 mV, 0.28 K). (b) Spectra in the BCS-BEC crossover regime (-5 mV, 0.1 nA, $V_{\rm{osc}}$ = 0.05 mV, 0.28 K). (c) Spectra in the BEC regime (-10 mV, 0.1 nA, $V_{\rm{osc}}$ = 0.1 mV, 0.28 K). (d) Zeeman splitting of the coherence peak in BEC regime as function of magnetic field. The blue dots are acquired by Gaussian fit. 
}
\end{figure} 

In experiments, a magnetic field up to 15 T is applied perpendicular to the FeSe monolayer. In the BCS regime, the superconducting gap and the coherence peaks are gradually suppressed with increasing magnetic field and completely vanish at 10 T [Fig. 4(a)]. The Zeeman splitting of the coherence peaks is visible. The Zeeman effect is more distinct in the crossover and BEC regimes [Figs. 4(b) and 4(c)]. The energy splitting of the coherence peak in Fig. 4(c) linearly depends on the magnetic field [Fig. 4(d)]. A g-factor of 2.0, which is close to that of a free electron, is extracted by linear fitting (see detail in \cite{SM}). These results are comparable to the calculation shown in Fig. 3(e). More investigation is needed to explore the exotic phases under strong magnetic fields.

In summary, we have provided direct spectroscopic evidence for a real-space BCS-BEC crossover in a FeSe monolayer on trilayer graphene substrate. The work function of the graphene substrate tunes the hole band edge in the FeSe monolayer crossing the Fermi level. As the top of hole band is shifted to and across the Fermi level, the BCS-BEC crossover is realized. The properties of the condensate are characterized by the the local density of states which we measured using scanning tunneling spectroscopy. We see a clear transition from a symmetric gap structure with sharp coherence peaks and sloped background in the BCS regime, to an asymmetric step-like gap structure with subdued peaks at gap edges. It is revealed that a pairing gap can even be formed in the incipient hole band when it is completely shifted below the Fermi level\cite{Bang2014,Ding15,Liu19}. The experimental data is well reproduced by a theoretical calculation using a two-band model. Finally, we have also mapped the behavior of local density of states in the presence of an applied magnetic field. Our experiment results show that FeSe monolayer is an ideal platform to study BCS-BEC crossover and possibly more exotic states under magnetic field.

\begin{acknowledgments}
We thank K. Chang, Q. Chen and C.-J. Wu for helpful discussions. This work is supported by the Ministry of Science and Technology of China (grant 2018YFA0305603) and the National Natural Science Foundation of China (grants 12074211, 11934001 and 51788104). S.K. gratefully acknowledges support from DFG KE-807/22-1.
\end{acknowledgments}

%

\end{document}